\documentclass[twoside,twocolumn,9pt]{article}
\usepackage{amsmath,amssymb}
\usepackage{extsizes}
\usepackage[super,sort&compress,comma]{natbib} 
\usepackage[version=3]{mhchem}

\usepackage[left=1.5cm, right=1.5cm, top=1.785cm, bottom=2.0cm]{geometry}
\usepackage{balance}
\usepackage{mathptmx}
\usepackage{sectsty}
\usepackage{graphicx} 
\usepackage{lastpage}
\usepackage[format=plain,justification=justified,singlelinecheck=false,font={stretch=1.125,small,sf},labelfont=bf,labelsep=space]{caption}
\usepackage{float}
\usepackage{fancyhdr}
\usepackage{fnpos}
\usepackage[english]{babel}
\addto{\captionsenglish}{%
  
}
\usepackage{array}
\usepackage{droidsans}
\usepackage{charter}
\usepackage[T1]{fontenc}
\usepackage[usenames,dvipsnames]{xcolor}
\usepackage{setspace}
\usepackage[compact]{titlesec}
\usepackage{hyperref}

\usepackage{epstopdf}

\definecolor{cream}{RGB}{222,217,201}

\begin{document}

\pagestyle{fancy}
\thispagestyle{plain}
\fancypagestyle{plain}{
\renewcommand{\headrulewidth}{0pt}
}

\makeFNbottom
\makeatletter
\renewcommand\LARGE{\@setfontsize\LARGE{15pt}{17}}
\renewcommand\Large{\@setfontsize\Large{12pt}{14}}
\renewcommand\large{\@setfontsize\large{10pt}{12}}
\renewcommand\footnotesize{\@setfontsize\footnotesize{7pt}{10}}
\makeatother

\renewcommand{\thefootnote}{\fnsymbol{footnote}}
\renewcommand\footnoterule{\vspace*{1pt}%
\color{cream}\hrule width 3.5in height 0.4pt \color{black}\vspace*{5pt}} 
\setcounter{secnumdepth}{5}

\makeatletter 
\renewcommand\@biblabel[1]{#1}            
\renewcommand\@makefntext[1]%
{\noindent\makebox[0pt][r]{\@thefnmark\,}#1}
\makeatother 
\renewcommand{\figurename}{\small{Fig.}~}
\sectionfont{\sffamily\Large}
\subsectionfont{\normalsize}
\subsubsectionfont{\bf}
\setstretch{1.125} 
\setlength{\skip\footins}{0.8cm}
\setlength{\footnotesep}{0.25cm}
\setlength{\jot}{10pt}
\titlespacing*{\section}{0pt}{4pt}{4pt}
\titlespacing*{\subsection}{0pt}{15pt}{1pt}

\fancyfoot{}
\fancyfoot[LO,RE]{\vspace{-7.1pt}\includegraphics[height=9pt]{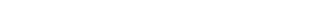}}
\fancyfoot[CO]{\vspace{-7.1pt}\hspace{13.2cm}\includegraphics{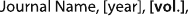}}
\fancyfoot[CE]{\vspace{-7.2pt}\hspace{-14.2cm}\includegraphics{head_foot/RF}}
\fancyfoot[RO]{\footnotesize{\sffamily{1--\pageref{LastPage} ~\textbar  \hspace{2pt}\thepage}}}
\fancyfoot[LE]{\footnotesize{\sffamily{\thepage~\textbar\hspace{3.45cm} 1--\pageref{LastPage}}}}
\fancyhead{}
\renewcommand{\headrulewidth}{0pt} 
\renewcommand{\footrulewidth}{0pt}
\setlength{\arrayrulewidth}{1pt}
\setlength{\columnsep}{6.5mm}
\setlength\bibsep{1pt}

\makeatletter 
\newlength{\figrulesep} 
\setlength{\figrulesep}{0.5\textfloatsep} 

\newcommand{\topfigrule}{\vspace*{-1pt}%
\noindent{\color{cream}\rule[-\figrulesep]{\columnwidth}{1.5pt}} }

\newcommand{\botfigrule}{\vspace*{-2pt}%
\noindent{\color{cream}\rule[\figrulesep]{\columnwidth}{1.5pt}} }

\newcommand{\dblfigrule}{\vspace*{-1pt}%
\noindent{\color{cream}\rule[-\figrulesep]{\textwidth}{1.5pt}} }

\makeatother
\twocolumn[
\begin{@twocolumnfalse}
{\includegraphics[height=30pt]{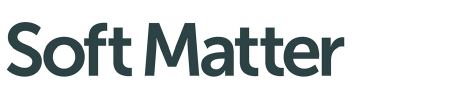}\hfill\raisebox{0pt}[0pt][0pt]{\includegraphics[height=55pt]{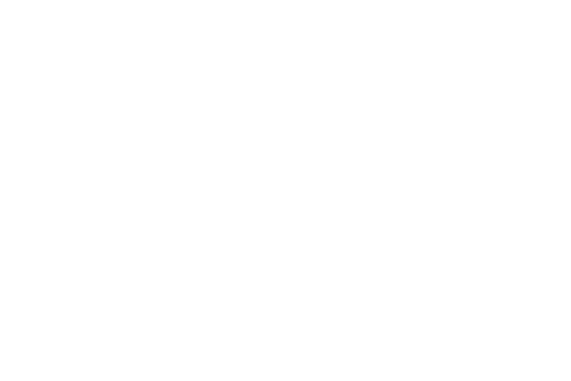}}\\[1ex]
\includegraphics[width=18.5cm]{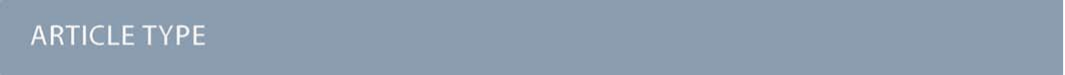}}\par
\vspace{1em}
\sffamily
\begin{tabular}{m{4.5cm} p{13.5cm} }

\includegraphics{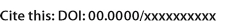} & \noindent\LARGE{\textbf{Dynamics of evaporating, interconnected droplets$^\dag$}} \\
\vspace{0.3cm} & \vspace{0.3cm} \\

 & \noindent\large{Chenyang Ren, \textit{$^{ab}$} Sri Ganesh Subramanian,\textit{$^{ab}$} Shresht Jain,\textit{$^{ab}$} Andrew Hazel,\textit{$^{bc}$} Finn Box,\textit{$^{ab}$} and  Anne Juel\textit{$^{\ast ab}$}    }   \\

\includegraphics{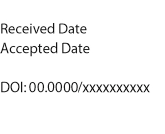} & \noindent\normalsize{We report on the dynamics of a pair of sessile droplets that are connected by a microchannel, yet open to the atmosphere and hence free to evaporate. 
Our results reveal that 
fluid exchange between droplets occurs via a pumping flow driven by differences in hydrostatic and Laplace pressure between the two droplets. 
Evaporation causes the droplets to slowly lose volume and change shape, which subsequently affects the fluid transport between them. 
We observe that, for equal contact areas, a larger droplet typically feeds a smaller droplet during evaporation and the flow in the connecting channel is unidirectional. However, for unequal contact areas, the flow can reverse in the connecting channel following a sudden switch in droplet shape that occurs during evaporation.
A stability analysis reveals that the dynamics of the exchange flow are underpinned by a supercritical pitchfork bifurcation. Evaporative volume loss permits the droplet pair to step through a sequence of quasi-stationary states determined by the instantaneous volume of the system. Enforcing unequal contact area unfolds the bifurcation such that droplet-shape switching and the associated flow reversal can be understood in terms of a jump from the disconnected to the connected branch of the bifurcation.
This establishes symmetry breaking as a mechanism to induce evaporation-driven flow reversal in connected droplets. } \\

\end{tabular}

\end{@twocolumnfalse} \vspace{0.6cm}

  ]

\renewcommand*\rmdefault{bch}\normalfont\upshape
\rmfamily
\section*{}
\vspace{-1cm}
\begin{@twocolumnfalse}
\end{@twocolumnfalse}

\footnotetext{\textit{$^{a}$~Manchester Centre for Nonlinear Dynamics, The University of Manchester, Manchester, M13 9PL, UK. E-mail: anne.juel@manchester.ac.uk}}
\footnotetext{\textit{$^{b}$~Department of Physics and Astronomy, The University of Manchester, Manchester,
M13 9PL, UK }}
\footnotetext{\textit{$^{c}$~Department of Mathematics, The University of Manchester, Manchester, M13 9PL,
UK }}

\footnotetext{\dag~Electronic supplementary information (ESI) available. See DOI: 10.1039/cXsm00000x/}


---------------------------

\section{Introduction}
\label{sec:intro}


Microfluidic devices that are passively driven, rather than requiring external pumping, have great potential as point-of-care diagnostic tools \cite{narayanamurthy2020advances}, due to their compactness and given that they do not require tethering to hardware, e.g., a pressure controller or syringe pump. 
Capillary action is one mechanism that provides a means of passively transporting fluids through stand-alone devices and hence rendering them portable. Capillary pumping, based on gradients in Laplace pressure, can be achieved in closed systems with `fluid walls' \cite{Walsh2017,nebuloni2023flows}, by confining liquid circuitry beneath an immiscible fluid layer to prevent evaporation of the working fluid. However, open microfluidics have the advantage of being accessible and straight-forward to manufacture \cite{de2016surface,Berry2019,berthier2019open} and have found application as diagnostic tools and in biochemical synthesis \cite{mcpherson2010microfluidic,de2016surface}. 

Droplet driven capillary micropumps \cite{walker2002passive,berthier2007flow,chen2009computation,xing2011droplet,moradi2023capillary} rely on flows between droplets of different sizes i.e., a `pumping droplet' and a `reservoir droplet', and flow control can be engineered by considering the material properties (hydrophobic vs.~hydrophilic) and geometry of the device and the droplets. 
On short time scales, droplet evaporation can typically be neglected. However, over longer time scales and in open systems, evaporation can significantly alter droplet shape, leading to changes in the pressure difference between the droplets, which in turn affects the pumping flow. Evaporation itself can be used to induce sustained microfluidic flows which can persist for minutes to hours \cite{lynn2009passive, chen2019capillary}, but its influence on open droplet-based microfluidic flows -- including the potential loss of working fluid -- is typically overlooked. 
Here, we investigate a model capillary micropump based on two connected droplets and simultaneously examine both the fluid transport between two droplets and the volume loss due to evaporation.

Our results reveal that, under certain conditions, the direction of capillary flow between droplets can reverse, rendering ephemeral the distinction between the `pumping' and `reservoir' droplets. Such a flow reversal has been previously observed in droplet-based micropumps \cite{ju2008backward, javadi2017flow}, but was attributed to inertial effects.  We model the dynamics of interconnected droplets in the absence of inertia and, in doing so, demonstrate that evaporation-induced flow reversal can instead be a direct consequence of device and droplet geometry.



\begin{figure*}[h!]
    \centering
    \includegraphics[width = 0.9\textwidth,keepaspectratio]{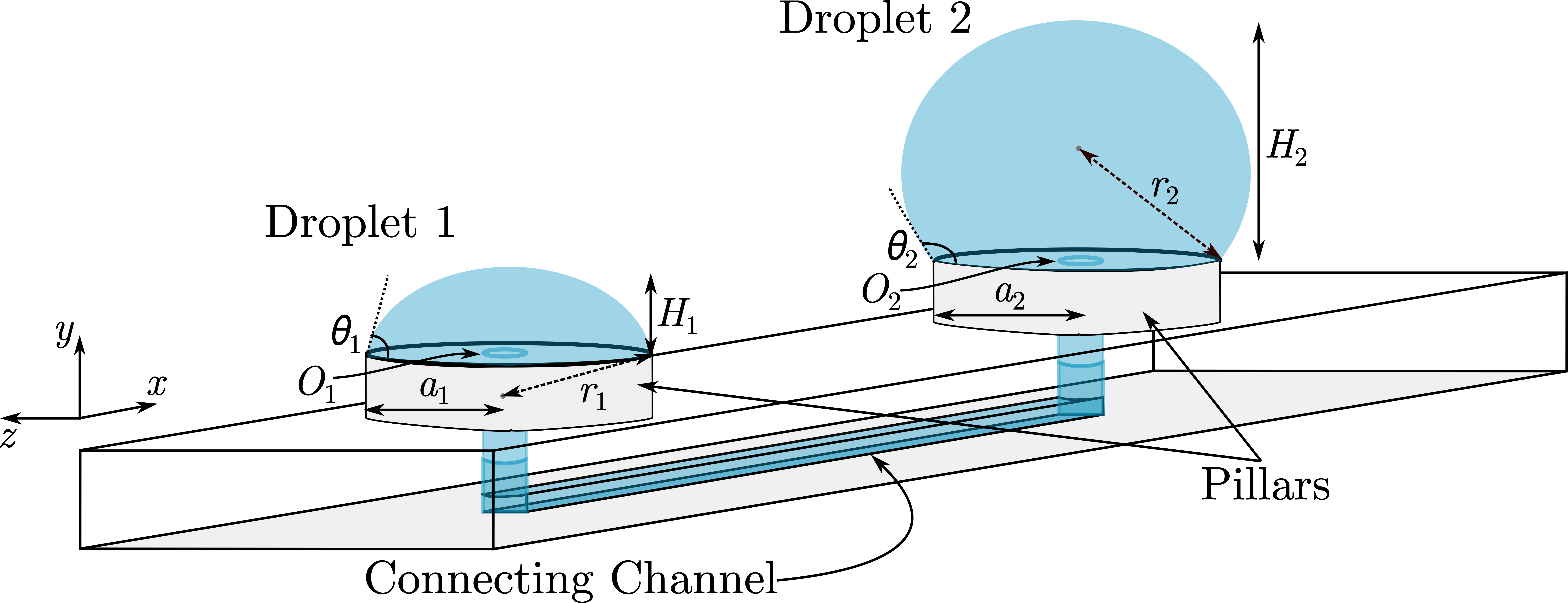}
    \caption{\textbf{Experimental system}. Schematic diagram of the experiment, comprising two sessile droplets resting on pillars of circular cross-section with radii $a_1$ and $a_2$, respectively. Each pillar has a circular hole at its center, denoted by $O_1$ and $O_2$, that forms the opening for a vertical, cylindrical channel. The two cylindrical channels are connected by a horizontal microchannel of rectangular cross-section. The contact angles of the droplets are represented by $\theta_1$ and $\theta_2$, while their heights are indicated by $H_1$ and $H_2$. Since the droplets resemble spherical caps, their radii of curvature, $r_1$ and $r_2$, were determined by fitting circles to the droplet profiles. Consequently, for a sessile droplet with a contact angle less than $\pi/2$, the center of the circular profile is depicted below the droplet base, whereas for a droplet with a contact angle above $\pi/2$, the center is located above the base. }
    \label{fig:expschem}
\end{figure*}


\section{Experimental method}
\label{sec:method}


In experiments, a pair of droplets was positioned upon two pillars connected via a channel. As the droplets evaporated, the evolving profile of each droplet was monitored and the flow field in the connecting channel was measured. A schematic diagram of the experimental set up is shown in figure~\ref{fig:expschem}.
Experiments were performed using droplets of deionized water, containing 0.04 $\%$ (w/v) polystyrene particles of diameter 5\,$\,\mu$m (TSI Inc KOBO, SP-500, refractive index 1.53, density $\rho=1.01\,$g/mL), in ambient laboratory environments for which the temperature and humidity were measured to be 23$\pm$0.1\textdegree\,C and 47$\pm$0.5 $\%$, respectively. The surface tension of the working fluid was measured to be $\gamma = 71.5\pm0.5\times10^{-3}$\,N/m using a Goniometer (Attension Theta Flex, Biolin Scientific), while we use literature values for viscosity $\mu = 0.95$\,mPas and density $\rho=1000\,$kgm$^{-3}$. 

The pillars and connecting channel were fabricated from Perspex using a micromiller. Prior to experimentation, the fabricated device was soaked in deionized water with 10$\%$ of a surface active cleaning agent (Decon 90) for 10 hours, and then rinsed in deionized water and dried with compressed nitrogen. This cleaning procedure prevented surface contamination -- a result of the micromilling process -- from affecting the surface tension of the droplets \cite{Ren2025}. 
Before droplets were deposited on the pillars, the connecting channel was filled manually using a syringe. The droplets were then carefully deposited upon their respective pillars using a micropipette (Acura 826 XS, Socorex) with range 0.5-10\,$\mu$L and accuracy 0.02\,$\%$-1.8\,$\%$. The contact line of each droplet remained pinned at the pillar edge throughout the experiment, such that droplet contact areas are constant. 
The diameter of pillars and initial droplet volumes were varied in experiments but the dimensions of the connecting channel remained the same; a horizontal section of length $10$\,mm and cross-section of $0.5$\,mm $\times$ $0.5$\,mm was connected at its ends to the base of each pillar via vertical cylindrical channels (see SI for exact channel geometry). 
The length of the connecting channel maintained the two droplets sufficiently far away from one another (i.e., greater than 10 droplet radii apart) that collective evaporative effects could be neglected \cite{Wray2020}. 
Droplets were illuminated with uniform, white back-lighting and side-view profiles were imaged using a CCD camera (Genie, Teledyne Dalsa) with a spatial resolution of 100\,pixels/mm and a frame rate of 40 images per second.

Throughout experiments, the Bond number $Bo = \rho g H^2/\gamma$, where $g$ is the acceleration of gravity and $H$ is the droplet height, is less than $0.86$, which implies that hydrostatic contributions to the pumping flow between droplets were small yet non-negligible relative to Laplace pressure contributions. However, droplet shape remained well-approximated by a spherical cap, so we acquired measurements of the radius, $R(t)$, height, $H(t)$ and volume $V(t)$ of each droplet during evaporation by fitting circles to droplet profiles and assuming axisymmetry; fits and measured profiles agreed to within $2.4\%$ throughout.

The pressure difference between channel ends were so small, however, that they were beyond the lower resolution limit of commercial pressure sensors. Hence, rather than measuring pressure differences, we assessed the pumping flow in the connecting channel using flow visualization.
To measure flow fields and flow velocities, the motion of tracer particles in the interconnecting channel was imaged using a microscope (AD4113ZTL, Dino-Lite) and we performed particle image velocimetry (PIV) on images acquired with a time interval of $1$\,s, using the PIVlab toolkit in Matlab. 
To measure the in-plane flow field in the connecting channel, the camera was placed underneath the channel, focused on the centre-line of the channel, and imaged from below with 130.5$\times$ magnification and a spatial resolution of 1.10\,$\mu$m/pixel. 
For modelling purposes, we are primarily concerned with the flow rate $Q=\iint_{S} u dS$ inside the connecting channel, where $u$ is the axial velocity and $S$ the cross-sectional area. We approximate $Q \approx \bar{u} S$ where $\bar{u}$ is the depth-averaged velocity that, for a channel of square cross-section, is related to the maximum speed along the center line by $u_{max}(z=0,y=0)=2.115\bar{u}$ 
\cite{munson1990fundamentals}.  
To visualize the flow inside the droplets, the mid-plane of the droplets was illuminated using a light sheet of width 200\,$\mu$m. To minimize the influence of heat on the evaporation rate of the droplet, we generated the light sheet using a cold light source (KL 2500 LED, Schott) in combination with a cylindrical lens.


\section{Theory}
\label{sec:theory}

The pressure at the channel ends, points $O_i$ in figure~\ref{fig:expschem}, comprises hydrostatic and Laplace pressure contributions, 
\begin{equation}
    P_i=P_{atm} +\frac{2\gamma}{r_i}+\rho g H_i,
\end{equation}
where $P_{atm}$ is atmospheric pressure, $\gamma$ is the surface tension of water, $H$ is the height of droplet and $r$ the radius of curvature of the free surface of a drop, $g$ is gravitational acceleration 
and $i=1,2$ denotes the droplet number. (We take the convention that $i=1$ represents the droplet on the left-hand side of side-view images and $i=2$ represents the droplet on the right). 

Fluid is exchanged between interconnected droplets via a pumping flow in the connecting microchannel. We assume the liquid to be both incompressible and Newtonian, and model the flow in the microchannel as fully-developed laminar flow with no-slip boundary conditions, 
such that the flow rate  $Q$ is related to the pressure drop between the two droplets by the Hagen-Poiseuille equation: 
\begin{equation}
    Q=K^{-1}\Delta P,
\end{equation}
where $\Delta P$ the pressure difference at the two ends of the connecting channel and $K=8.73\times 10^9$\,kgm$^{-4}$s$^{-1}$ is the resistance coefficient of the connecting channel (see SI for further details).

Droplet evaporation is controlled by the diffusion of vapour from the free surface to the ambient environment. 
The rate of volume loss of an individual sessile droplet with constant contact radius is given by \cite{wilson2023evaporation}
 \begin{equation}
    \frac{dV_{{evap},i}}{dt} 
    = \pi B_i a_i^3 \frac{f(\theta_i)}{\mathrm{sin}\theta_i},
\end{equation}
where
\begin{equation}
    B_i=\frac{-2D \Delta p M}{\rho R T {a_i}^2}. 
\end{equation} 

The volume of each droplet, $V_{evap, i}$, evolves in time because of evaporation, which is driven by a pressure difference $\Delta p=p_0-p_\infty$ between the saturation vapour pressure of water, $p_0=2.065$\,kPa, and ambient vapour pressure far away from the surface of the droplets $p_\infty$; here, we determined the ambient vapour pressure from the measured relative humidity $RH = p_\infty/p_0 = 0.47$, giving $\Delta p = p_0(1-RH) = 1.094$\,kPa. 
Since the contact radius of the droplets, $a_i$, is fixed by the pillar geometry, the apparent contact angle of the droplets, $\theta_i$, evolves in time. Here, $D = 2.3 \times 10^{-5}$\, m$^2$\,s$^{-1}$ is the diffusion coefficient of water vapour molecules in air, $M=18$\,g/mol is the molar mass of water, and $R=8.3$\,J\,mol$^{-1}$K$^{-1}$ and $T$ are the gas constant and ambient temperature, respectively, and $f(\theta )$ is an empirical function \cite{PICKNETT1977,SCHONFELD2008} given by 
\begin{equation}
f(\theta) = 0.00008957 + 0.6333\theta + 0.116\theta^2 -0.08878\theta^3 + 0.01033\theta^4.
\end{equation}

For a pair of interconnected droplets, the volume of each individual droplet is therefore influenced both by evaporation and the pumping flow within the connecting channel. As such, the evolution of individual droplet volumes is coupled, and can be written in terms of volumetric change associated with both evaporation and pumping flow,
\begin{equation}
\begin{pmatrix}
\frac{dV_1}{dt}\\\\
\frac{dV_2}{dt}
\end{pmatrix} = \begin{pmatrix} 
\frac{dV_{{evap},1}}{dt}-K^{-1}(P_1- P_2)\\\\
\frac{dV_{{evap},2}}{dt}+K^{-1}(P_1- P_2)
\end{pmatrix}.
\label{eq:evo}
\end{equation}

Assuming that the droplets take the form of a spherical cap we invoke the following geometrical relations between droplet volume $V_i$, height $H_i$, radius $r_i$ and contact angle $\theta_i$:
\begin{equation}
  V_i=\frac{\pi a_i^3 (2-3\cos\theta_i+\cos^3 \theta_i)}{3\sin ^3\theta_i},
\end{equation}   
\begin{equation}
  H_i=\frac{ a_i(1-\cos\theta_i)}{\sin \theta_i},
\end{equation}   
\begin{equation}
 r_i=\frac{ a_i}{\sin \theta_i}.
\end{equation}  

This permits to rewrite eq.~\ref{eq:evo} in terms of droplet contact angles, $\theta_i$, as
 \begin{equation}
        \begin{pmatrix}
         \Dot{\theta_1}\\\\
         \Dot{\theta_2}
        \end{pmatrix} = \begin{pmatrix}
         B_1 \frac{f(\theta_1)(\cos{\theta_1}+1)^2}{\sin{\theta_1}} - K^{-1}(P_1(\theta_1)-P_2(\theta_2)) \frac{(\cos{\theta_1}+1)^2}{\pi a_1^3}\\\\
         B_2 \frac{f(\theta_2)(\cos{\theta_2}+1)^2}{\sin{\theta_2}} + K^{-1}(P_1(\theta_1)-P_2(\theta_2))\frac{(\cos{\theta_2}+1)^2}{\pi a_2^3}
        \end{pmatrix},  
        \label{eq:angle_evo}
    \end{equation}

where
\begin{equation}
    P_i(\theta_i)=\frac{2\gamma\sin \theta_i}{a_i}+ \frac{\rho g a_i(1-\cos\theta_i)}{\sin\theta_i}.
    \label{eq:drop_pressure}
\end{equation}

Since the evaporative terms are dissipative, eq.~(\ref{eq:angle_evo}) has one stationary solution, $\theta_1=\theta_2=0$, which physically implies that evaporation can drive a pair of interconnected droplets to evolve to have net zero volume, $V_1=V_2=0$, irrespective of initial conditions. The evolution of $\theta_i$ with time can be calculated numerically with the built-in initial value problem solver ODE45 in Matlab, for initial values of $\theta_{i,t=0}$.


\section{Results}
\label{sec:results}
\subsection{Experimental Observations}
\label{subsec:expresults}

\begin{figure*}[h!]
    \centering
    \includegraphics[width =1\textwidth,keepaspectratio]{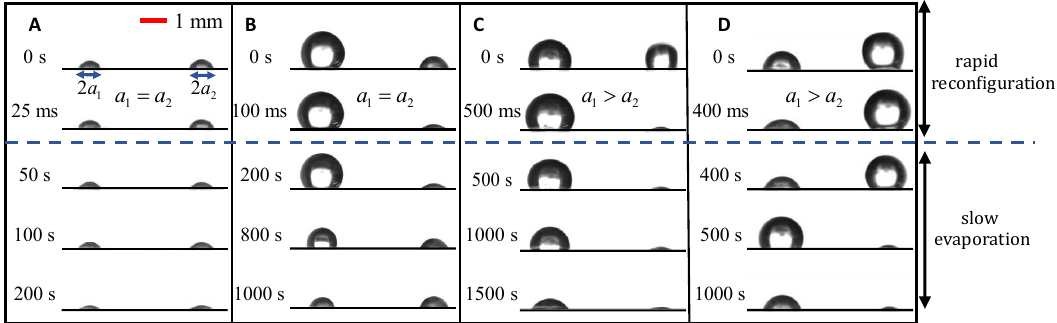}
    \caption{\textbf{Evolution of a pair of connected evaporating droplets}. Images acquired at various instance in time (as indicated) demonstrate a rapid redistribution of fluid between droplets at early times ($t\lesssim 1$\,s) followed by a slow evolution of droplet shape at later times ($t\gg 1$\,s). A. Two droplets of low initial volume on connected pillars of equal radii adopt a wetting configuration, such that both droplets have initial contact angles $\theta_1, \theta_2<\pi/2$. They rapidly acquire the same volume  ($t\lesssim 25$\,ms) and then remain approximately identical in size and shape as they evaporate. B. Two droplets of markedly-different initial volume on connected pillars of equal radii adopt an asymmetric configuration with $\theta_1 > \pi/2$ and $\theta_{2}<\pi/2$. As they evaporate, the volume of the larger droplet decreases while the volume of the smaller droplet increases until the two have the same volume, which indicates that there is a net flow in the connecting channel from the bigger drop to the smaller drop. C. Droplets on unequal pillars; larger droplet on the larger pillar. After the rapid transition period ($t\lesssim 500$ ms), $\theta_1 > \pi/2$ and $\theta_{2}<\pi/2$. With evaporation, the volume of the smaller droplet remains nearly constant, indicating that there is a net flow in the connecting channel from the bigger droplet to the smaller droplet.
    D. Droplets on unequal pillars; smaller droplet on the larger pillar. 
    During the slow evaporation stage the droplet volumes switch suddenly; at $t\sim400$\,s the position of the largest drop rapidly changes from pillar $2\rightarrow 1$). }
    \label{fig:exp-images}
\end{figure*}

Our experiments reveal that the evolution of an interconnected droplet pair occurs on two timescales, see figure~\ref{fig:exp-images}, as fluid is exchanged through the connecting channel and evaporation occurs. 
The influence of the pumping flow is most clearly apparent at early times, evidenced by a rapid redistribution of fluid between droplets that causes sudden changes in the size of individuals droplets, within $t\lesssim 1$\,s of being deposited on the pillars. We note that the time taken for this initial rapid reconfiguration to occur varies between experiments because of differences in initial droplet volumes. On a longer timescale, $t\gg 1$\,s, droplet volumes change slowly as evaporation decreases the total volume of the system. However, a pumping flow may persist during the slow evaporation stage, and act to replenish fluid lost through evaporation such that a single droplet may maintain its size as the volume of the other droplet decreases.

\begin{figure*}[h!]
    \centering
    \includegraphics[width = 0.85\textwidth,keepaspectratio]{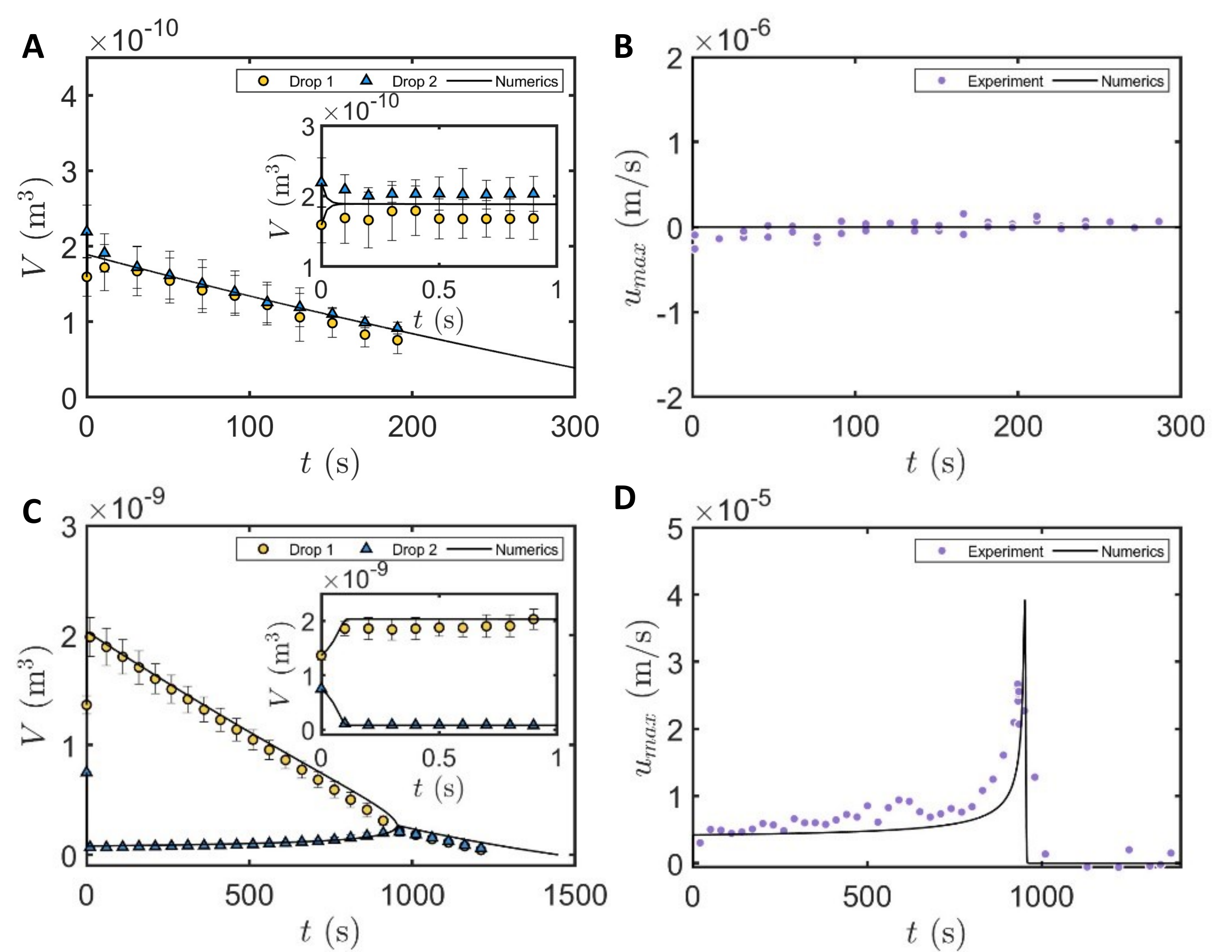}
    \caption{\textbf{Connected droplets on equal pillars.} Evolution of droplet volumes (A,C) and channel flow speed (B,D) measured as a function of time $t$ for pillars of equal radii $a_1=a_2=0.5$ mm. A-B: $V_{1,t=0}=0.15$ $\mu$L and $V_{2,t=0}=0.25$ $\mu$L; the droplets are comparable in size during evaporation and the pumping flow between the droplets is measured to be approximately zero throughout.  C-D. $V_{1,t=0}=1.45$ $\mu$L and $V_{2,t=0}=0.8$ $\mu$L; the difference in droplet size generates a net pumping flow from drop 1 to drop 2 (i.e., from the bigger drop to the smaller drop) during evaporation. Experimental data, from figure~\ref{fig:exp-images}A-B, represented by markers (see legend), numerical data represented by black lines. Insets: zooming in on early times ($t\lesssim 1$) highlights the rapid redistribution of fluid between two droplets that occurs immediately after they are connected.   } 
    \label{fig:equal-pillars}
\end{figure*}

First, we consider the case of droplets on connecting pillars of equal radii i.e., for $a_1 = a_2 = 0.5$\,mm, see figure~\ref{fig:exp-images}A-B. 
When both droplets have initial contact angles $\theta_1, \theta_2<\pi/2$, they rapidly acquire the same volume, as shown in figure~\ref{fig:exp-images}A for which $V_{1,t=0}=0.15$\,$\mu$L and $V_{2,t=0}=0.25$\,$\mu$L.  
Initially,  within $t\lesssim 25$\,ms of deposition, both the hydrostatic and Laplace contributions to pressure combine to drive fluid from the bigger droplet to the smaller droplet, see inset to figure~\ref{fig:equal-pillars}A. As they then slowly evaporate on a longer timescale, $t\gg 1$\,s, the two droplets remain approximately identical in size and shape, see figure~\ref{fig:equal-pillars}A. As such, the pressure differential across the channel is null, and the evaporative flux of each droplet is equivalent, so the mean flow velocity within the connecting channel is zero, see figure~\ref{fig:equal-pillars}B. 

If instead, 
the contact angle of at least one droplet is $\theta_i>\pi/2$, then at early times the smaller droplet shrinks and the larger droplet grows 
so that an asymmetric configuration is rapidly established with notable differences in droplet volume and contact angle between the drops. For $V_{1,t=0}= 1.45 $ $\mu$L and $V_{2,t=0}= 0.8 $\,$\mu$L this rapid redistribution of fluid occurs within ($t\lesssim 100$\,ms), as shown in figure~\ref{fig:exp-images}B, and suggests that the pumping flow is initially dominated by a Laplace pressure differential. However, as the drops evaporate, the larger droplet shrinks while the smaller droplet grows slightly until the droplets match in size and evaporate out in unison, see figure~\ref{fig:equal-pillars}C. 
Although a flow driven by differences in evaporative flux might be expected to wick fluid towards the larger droplet, which has greater surface area and thus larger evaporative flux,
our measured flow velocities in the channel instead provide evidence of a unidirectional flow from the larger droplet to the smaller droplet, see figure~\ref{fig:equal-pillars}D, 
implying that the pumping flow is instead governed by a pressure differential based on capillary statics i.e.,  Laplace and hydrostatic contributions. 
Our measurements show 
that the flow velocity diverges as the droplet volumes become approximately equal, after which it reduces to zero. This implies that pressure differences between the drops continue to drive the flow in the connecting channel as evaporation causes the droplets to change shape. 

The distinction between these two aforementioned cases arises from the nonlinear pressure-contact angle relationship of a single sessile droplet, given by eq.~(\ref{eq:drop_pressure}), which has a local maxima for $\theta \approx \pi/2$ (see SI and e.g., ref.~\cite{xing2011droplet} for further details) that corresponds to a hemispherical drop i.e., for $V_i \approx 2\pi a_i^3/3$.
A pair of droplets, each with maximal pressure, would therefore have a total volume equal to that of a sphere with an equatorial cross-section given by the contact area (i.e., pillar base) and this serves as a threshold value, $V_{\mathrm{tot}}^{\mathrm{crit}} \approx 4\pi a^3/3$ where $a = a_1 = a_2$.
For a pair of connected droplets with equal contact area and $\theta_1, \theta_2<\pi/2$, such that $V_{\mathrm{tot}} \lesssim V_{\mathrm{tot}}^{\mathrm{crit}}$, instantaneous pressure equalization can only be achieved with equal volume drops, since the the pressure-volume relation of a single drop is monotonic for $\theta <\pi/2$. 
Instead, for $V_{\mathrm{tot}} \gtrsim V_{\mathrm{tot}}^{\mathrm{crit}}$ instantaneous equilibrium can be attained with either an unequal pair (e.g., $\theta_1<\pi/2$ and $\theta_2>\pi/2$), since contact angle is a double-valued function of pressure in the range $0<\theta<\pi$, 
or with an equal pair for which $\theta_1 = \theta_2>\pi/2$. 
In Section~\ref{sec:results}.C, we demonstrate that the latter case is an unstable configuration, however, such that perturbations will drive the pair to instead take on different volumes and contact angles.

\begin{figure*}[h!]
    \centering
    \includegraphics[width = 0.85\textwidth,keepaspectratio]{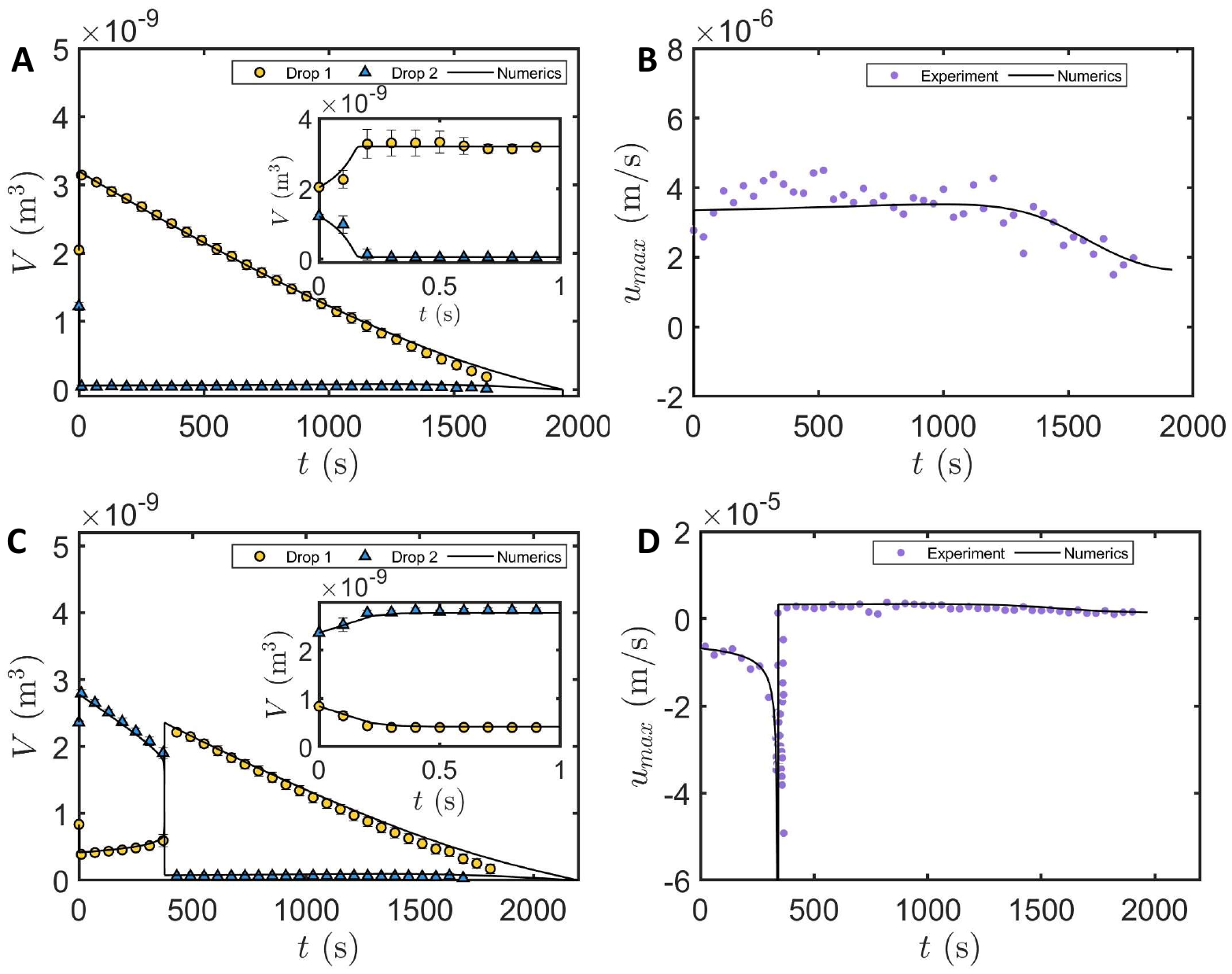}
    \caption{\textbf{Connected droplets on unequal pillars.} Evolution of droplet volumes (A,C) and channel flow speed (B,D) measured as a function of time $t$ for pillars of radii $a_1=0.75$ mm and $a_2=0.5$ mm, respectively. A-B: $V_{1,t=0}=2$\,$\mu$L and $V_{2,t=0}=1.2$\,$\mu$L; during evaporation, the difference in droplet size generates a net pumping flow from drop 1 to drop 2 (i.e., from the bigger drop to the smaller drop).  C-D. $V_{1,t=0}=0.8$\,$\mu$L and $V_{2,t=0}=2.3$\,$\mu$L; initially the bigger drop is positioned on the smaller pillar and the pumping flow is from drop 2 to drop 1 but, for $t\sim 400$\,s, this configuration inverts suddenly and flow reversal is observed in the connecting channel. Experimental data, figure~\ref{fig:exp-images}C-D, represented by markers (see legend), numerical data represented by black lines. Insets: zooming in on early times ($t\lesssim 1$) highlights the rapid redistribution of fluid between two droplets that occurs immediately after they are connected. } 
    \label{fig:unequal-pillars}
\end{figure*}

The interplay of droplet geometry and pressure contributions can be further disentangled by examining interconnected droplets on pillars of unequal radii i.e., for $a_1 > a_2$, see figure~\ref{fig:exp-images}C-D.  
First we consider the case in which both droplets have sufficient volume that their initial contact angles $\theta_1, \theta_2>\pi/2$, and the bigger droplet is positioned on the bigger pillar -- as shown in figure~\ref{fig:exp-images}C for which $V_{1,t=0}= 2$\,$\mu$L, $V_{2,t=0}= 1.2$ \,$\mu$L, $a_1=0.75$ mm and $a_2=0.5$ mm. Initially, fluid is rapidly redistributed from the smaller to the larger drop (see inset to figure~\ref{fig:unequal-pillars}A), suggesting that the Laplace pressure differential dominates the pumping flow at early times. After this rapid transition, an instantaneous pressure balance is reached with $\theta_1 > \pi/2$ and $\theta_{2}<\pi/2$. As evaporation occurs, and the system evolves, fluid flows from the larger to the smaller drop such that the larger droplet shrinks whilst the smaller droplet maintains its size, see figure~\ref{fig:unequal-pillars}A-B.

\begin{figure*}[h!]
    \centering
    \includegraphics[width = 0.95
    \textwidth,keepaspectratio]{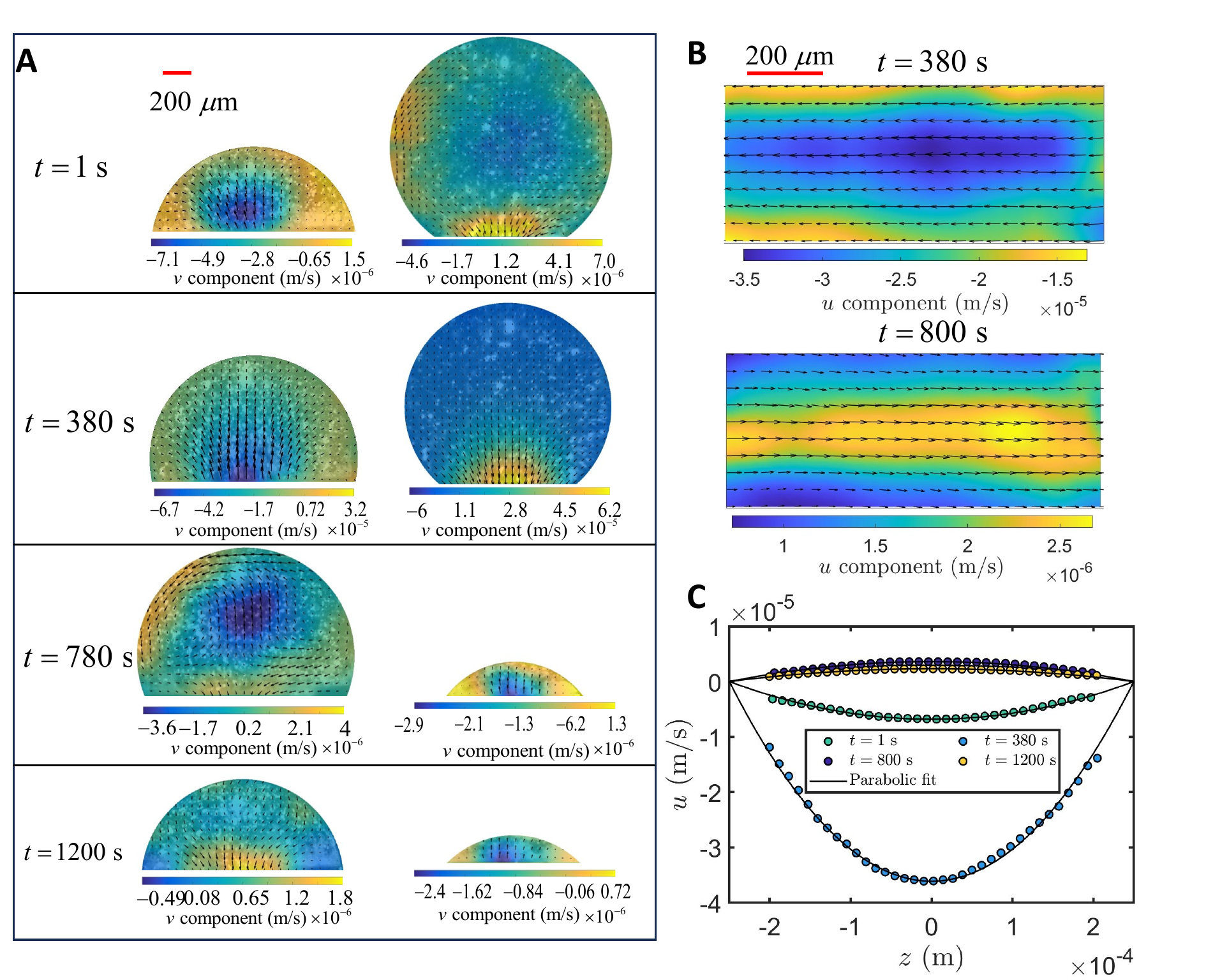}
    \caption{\textbf{Visualising flow reversal}. A. Flow fields, measured on the mid-plane, of two connected droplets at various instances in time following deposition, as indicated; the colorbar represents the magnitude of the velocity in the $y$-direction.
    The droplets' initial total volume $V_{\mathrm{tot}, t=0}= 3\,\mu$L and, as they evaporate, the droplets switch shape and the flows inside the droplets and in the connecting channel reverse direction. Droplet-shape switching and flow reversal occurs at $t\sim 400$\,s. 
    B. Instantaneous flow fields measured in the connecting channel before and after droplet switching demonstrate flow reversal; the colorbar represents the magnitude of the velocity in the $x$-direction.
    C. Measured velocity profiles within the connecting channel at various instances in time before and after droplet-shape switching, as indicated. Data in B-C is from the same experiments as that shown in figure~\ref{fig:unequal-pillars}C-D. Data in A is from a different experiment but with the same initial conditions. }
    \label{fig:dropflow}
\end{figure*}

In all of the cases presented so far, the pumping flow established during evaporation has been unidirectional (from the larger to the smaller droplet) or null (for equal volume droplets on equal radii pillars). However, if the smaller droplet, $\theta_1<\pi/2$, is initially positioned on the larger pillar and the larger droplet, $\theta_2>\pi/2$, on the smaller pillar, we instead see flow reversal during the slow evaporation stage that is associated with a sudden switching of droplet shapes, see figures~\ref{fig:exp-images}D and \ref{fig:unequal-pillars}C-D. 
Following droplet-shape switching, the pumping flow in the connecting channel continues to feed the smaller droplet as they evaporate out.
In figure~\ref{fig:dropflow}, we show measured flow fields inside the two droplets and the connecting channel, and velocity profiles within the connecting channel, at various instants during droplet evaporation, for the same initial conditions as the experiment shown in figure~\ref{fig:unequal-pillars}C-D.
Flow visualization demonstrates how droplet-shape switching and flow reversal combine to maintain the pumping flow from larger droplet to smaller droplet. 
Visualization of flow fields inside the droplets show that before droplet-shape switching, for $t\lesssim 380$\,s, the smaller droplet (droplet 1 in figure~\ref{fig:dropflow}A) is fed by the larger droplet (droplet 2 in figure~\ref{fig:dropflow}A) and likewise after switching, for $t\gtrsim 380$\,s, the larger droplet continues to feeds the smaller droplet. 
Flow velocities in the channel were acquired by spatial averaging in the $x$-direction of instantaneous flow fields, exhibit the parabolic profile typical of fully-developed Hagen-Poiseuille flow, and show how the flow in the connecting channel reverses direction for $t\gtrsim 380$\,s, see figure~\ref{fig:dropflow}C.
Using measured internal flow velocities, we estimate the scale of the pressure difference induced by the internal flow within the droplets to be $\Delta P_f \sim \mu (\Delta v/\Delta y) \sim 10^{-6}$\,Pa, where $\mu$ is the dynamic viscosity of fluids, and $\Delta v$ is the difference in 
velocity in the $y$-direction, between the droplet apex and the channel outlet, such that the distance over which the pressure difference occurs scales with droplet height $\Delta y \sim H_i$. 
Since this flow-induced contribution to the pressure is much smaller than either the Laplace or hydrostatic contributions i.e., on the order of $\mathcal{O}(10^{-6})$\,Pa compared to $\mathcal{O}(100)$\,Pa to $\mathcal{O}(10)$\,Pa, respectively, we neglect the pressure contributions due to internal droplet flows in our analysis. (Although we note that the inertia of this internal flow was responsible for flow reversal in \citet{ju2008backward} and \citet{javadi2017flow}).



\subsection{Droplet dynamics}

\begin{figure*}[h!]
    \centering
    \includegraphics[width = 1\textwidth,keepaspectratio]{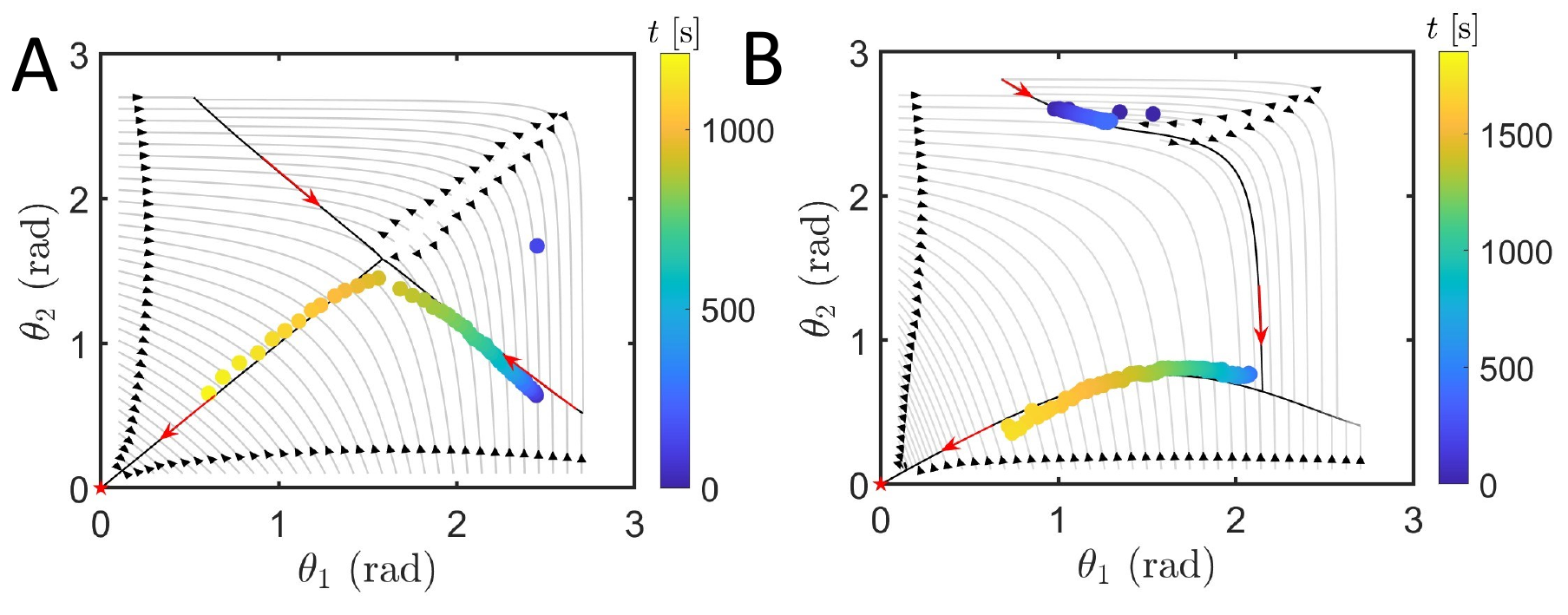}
    \caption{\textbf{Trajectories in pseudo-phase space}. Phase portrait of the evolutionary trajectories of an evaporating, interconnected droplet pair, in the pseudo-phase space of the their respective contact angles, for A. equal pillars of radii $a_1=a_2=0.5$\,mm, and B. unequal pillars of radii $a_1=0.75$\,mm and $a_2=0.5$\,mm. Numerical data of the evolution of the contact angles of the two droplets is represented by lines (faint, grey lines represent early-time behaviour for $t<1$\,s and black lines represent late-time behaviour for $t>1$\,s), with arrows indicating the direction of droplet evolution. The red star (at $\theta_1 = \theta_2 = 0)$ indicates the stationary solution of equation (13), which is the terminal point of the evaporating droplets.
    Experimental data (for the same parameters as indicated in Fig.~3C and 4C, respectively) is represented by markers where the colour indicates time $t$, as indicated in the legend colormap. The red arrows indicates the direction of the droplets evolution with time in the experiments. The black arrowheads are positioned at $t=10$~ms along trajectories.  
    }
    \label{fig:phase_space}
\end{figure*}

Numerical solutions to the contact angle evolution equation, eq.~\ref{eq:angle_evo}, enable us to construct a pseudo-phase portrait of an interconnected droplet pair; initial values of $\theta_i$ were chosen for portrait clarity. In figure~\ref{fig:phase_space}A-B, we present dynamical portraits of an interconnected droplet-pair in the phase space of the respective contact angles of the two droplets, $(\theta_1, \theta_2)$, for pillars of equal and unequal radii. 
Trajectories represent the time-evolution of an inter-connected droplet pair; as evaporation progresses and the total volume decreases, the contact angles of each droplet evolve until reaching the stationary point where $\theta_1=\theta_2=0$.
To highlight the two distinct timescales of the system, we plot solutions for $t<1$\,s as faint grey lines and solutions for $t>1$\,s as solid black lines, and include experimental data that is colour-coded by time after droplet deposition. 

For pillars of equal radii, $a_1=a_2=0.5$\,mm, we see that droplet states rapidly converge onto slow evaporation pathways, see figure~\ref{fig:phase_space}A. 
If both droplets are initially in a wetting configuration i.e., for $\theta_{1,t=0}<\pi/2$ and $\theta_{2,t=0}<\pi/2$, the system quickly converges onto a slow evaporation pathway defined by $\theta_1=\theta_2<\pi/2$, and remain on that path as the total volume of the system decreases until $\theta_1=\theta_2=0$. As the droplets have equal contact angles and equal contact areas, there exists zero pressure difference between channels ends in this scenario, and hence zero pumping flow along this path (droplets evaporate out together in unison). 
However, if either of the droplets has an initial contact angle $\theta_{i,t=0}>\pi/2$, and $V_{\mathrm{tot},t=0} \gtrsim 4\pi a_i^3/3$, then the system rapidly converges onto an evaporation path with one increasing contact angle and one decreasing contact angle, for which a pumping flow feeds the smaller droplet, before eventually reaching the low-volume evaporation pathway defined by $\theta_1=\theta_2<\pi/2$. Our numerical results also indicate trajectories rapidly diverge from the initial conditions $\theta_1=\theta_2>\pi/2$ which suggests this an unstable state; the stability of the system is discussed further in Section \ref{subsec:stability}.

In the case of unequal pillars, with $a_1=0.75$\,mm and $a_2=0.5$\,mm, the symmetry of the system is broken, and this is reflected in the resulting pseudo-phase portrait, see figure~\ref{fig:phase_space}B. We find that initial conditions rapidly converge onto one of two evaporative pathways. On the $\theta_2>\theta_1$ path, $\theta_2$ decreases while $\theta_1$ increases, as time progresses. On the $\theta_1>\theta_2$ path, $\theta_1$ decreases while $\theta_2$ initially increases and then decreases, as time evolves. However, our experiments (see figure~\ref{fig:exp-images}D) show that a droplet pair can instantaneously jump from one pathway to the other, which manifests in experiments as a droplet-shape switching event and flow reversal in the connecting channel. 


\subsection{Stability of an interconnected droplet pair}
\label{subsec:stability}

The typical timescales of fluid transport between droplets is much less than the evaporative lifetime of the droplets. Indeed, in the limit where evaporation is a much slower process than the pumping flow, 
we can consider the dynamics on the short timescale associated with fluid transportation between droplets, i.e., for a constant volume, 
and neglect the slow dynamics associated with evaporation. 
The instantaneous configuration of the droplet pair can thus be considered to be a quasi-steady state, corresponding to specific droplet volumes and a corresponding set of contact angles; 
as the droplets evaporate, the system steps through a sequence of quasi-steady states that are determined by the instantaneous total volume of the system. 

To assess the stability of different droplet-pair configurations for a given total volume, $V_\mathrm{tot} = V_1+V_2$, 
we note that, on the timescale of the pumping flow, the equations that govern the dynamics of the contact angle can be expressed as:
\begin{equation}
 \begin{pmatrix}
         \Dot{\theta_1}\\\\
         \Dot{\theta_2}
        \end{pmatrix} =  \begin{pmatrix}
    
        - (P(\theta_1,a_1,\gamma,\rho,g)-P(\theta_2,a_2,\gamma,\rho,g))\frac{(\cos{\theta_1}+1)^2}{\pi a_1^3} \\\\
          (P(\theta_1,a_1,\gamma,\rho,g)-P(\theta_2,a_2,\gamma,\rho,g))\frac{(\cos{\theta_2}+1)^2}{\pi a_2^3}
        \end{pmatrix}.
        \label{eq:angle_evo_short_time}
\end{equation}


Eq.~\ref{eq:angle_evo_short_time} has two sets of stationary solutions: 
one solution set is given by $P(\theta_1,a_1,\gamma,\rho,g)-P(\theta_2,a_2,\gamma,\rho,g) = 0$, which corresponds to a solution governed by the exchange of fluid between droplets; the other set of stationary solutions, $\theta_1=\theta_2=\pi$, represents spherical droplets, which is not relevant to our experiments. To determine the stability of stationary points, we perturb the relevant set of solutions and numerically calculate eigenvalues of the corresponding Jacobian matrix; stability is inferred from the sign of the least negative eigenvalue. \\

\begin{figure*}[h!]
    \centering
    \includegraphics[width = 0.95\textwidth,keepaspectratio]{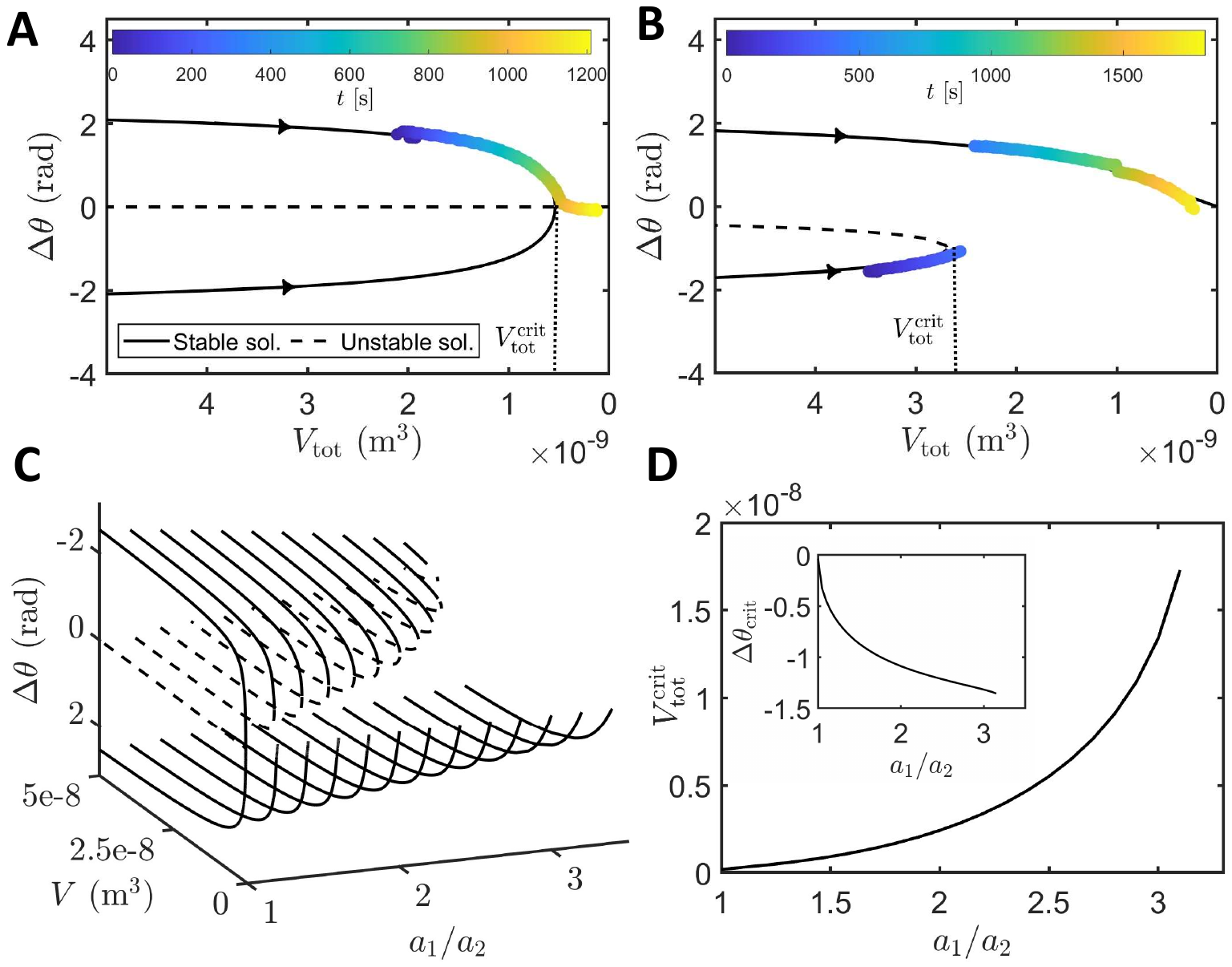}
    \caption{\textbf{Pseudo-stability of a pair of connected evaporating droplets.} Bifurcation diagrams of the difference in contact angle $\Delta \theta = \theta_1 - \theta_2$ measured as a function of the total volume $V_{\mathrm{tot}} = V_1 + V_2$ for A. equal pillars, $a_1=a_2=0.5$\,mm, manifests a pitchfork bifurcation,  B. unequal pillars, $a_1=0.75$ mm and $a_2=0.5$\,mm, manifests an imperfect pitchfork bifurcation. and C. varying ratio of pillar radii in the range $1<a_1/a_2<3.5$,  which demonstrates the unfolding of the pitchfork bifurcation for increasing $a_1/a_2$. Results of a numerical stability analysis of the solutions of governing equation without evaporation (eq.~\ref{eq:angle_evo_short_time}) are represented by black lines; solid lines indicate stable solutions, while dashed lines indicate unstable solutions. D. Volume $V_{\mathrm{tot}}^{\mathrm{crit}}$, and difference in contact angle $\Delta \theta_{\mathrm{crit}}$ (inset), at the limit-point calculated as a function of imperfection parameter $a_1/a_2$.
    In A-B, experimental data (for the same parameters as indicated in figures 3C-D and 4C-D, respectively) is represented by markers where the colour indicates time $t$, as indicated in the legend colormap. 
    }
    \label{fig:stability}
\end{figure*}

We present the findings of our linear stability analysis of droplet-pair configurations in a parameter space based on the difference in contact angle $\Delta \theta = \theta_1 - \theta_2$ and the total volume $V_{\mathrm{tot}} = V_1 + V_2$ in figure~\ref{fig:stability}. Stable solutions are represented by solid lines while the unstable solution is represented by the thick dashed line. 

In the case of pillars of equal radii, $a_1 = a_2$, we uncover a supercritical pitchfork bifurcation, see figure~\ref{fig:stability}A, that breaks droplet-shape symmetry. For low total volumes, one stable fixed point exists and corresponds to a symmetric droplet-pair configuration with $\Delta\theta = 0$. As the total volume increases, 
the system bifurcates from having one fixed point to three fixed points; the branching point, $V_\mathrm{tot}^\mathrm{crit} \approx 4\pi a ^3/3$, is a consequence of both droplets having constant contact area and hence a nonlinear pressure-volume relation with maxima for $\theta = \pi/2$ 
(as discussed in Section~\ref{sec:results}.A). 
In the triple-valued region, asymmetric configurations are stable, and correspond to an unequal droplet pair with pumping flow from the larger to the smaller droplet, while the symmetric solution becomes unstable. This implies that, as time evolves and volume decreases because of evaporation, droplets regain symmetry (i.e., smoothly transition from an asymmetric to a symmetric configuration) once $V_{\mathrm{tot}} \leq V_\mathrm{tot}^\mathrm{crit}$, before evaporating out together. 

Breaking the symmetry of the system, by using pillars of unequal radii, leads to an imperfect pitchfork bifurcation, see figure~\ref{fig:stability}B. The upper branch now transitions smoothly down to $V_\mathrm{tot} = 0$ while the stable lower branch meets the unstable symmetric solution at a finite $V_\mathrm{tot}^\mathrm{crit}$. As such, a droplet pair that begins on the disconnected, stable asymmetric branch must jump catastrophically onto the connected asymmetric branch once evaporation reduces the system volume below this critical value. Indeed, the droplet-shape switching and accompanying flow reversal, shown in figures~\ref{fig:exp-images}D, \ref{fig:unequal-pillars}D and \ref{fig:dropflow}, are manifestations of the discontinuous jump from one stable branch, that corresponds to the smaller droplet having larger contact area, to the other stable branch, that corresponds to the larger droplet having larger contact area, which occurs as evaporation reduces the total volume of the system. 

Our stability analysis shows the unfolding of this pitchfork bifurcation is governed by the ratio of the two pillar radii, $a_1/a_2$, see figure~\ref{fig:stability}C. Increasing this imperfection parameter causes the size of the disconnection to increase. Physically, in the range $1 < a_1/a_2  \leq \pi$, the limit point corresponds to the total volume of the droplet pair when the droplet with larger contact area is hemispherical, with maximal local pressure for $\theta_1 \approx \pi/2$, while the droplet with smaller contact area has a non-wetting configuration with a volume that supplies the same pressure, such that $V_\mathrm{tot}^\mathrm{crit} \approx 2\pi a_1/3 + V_2(P_2 = P_1(\theta_1 \approx \pi/2),\theta > \pi/2)$, as shown in figure~\ref{fig:stability}D. For $a_1/a_2>\pi$, only one stable branch exists and that corresponds to a single droplet residing on the larger pillar.


\section{Conclusions \& Discussion}

We have shown that the dynamics of an interconnected droplet pair is governed by two timescales; a short timescale associated with the pumping flow that exchanges fluid between the two drops, and is determined by the pressure differential across the connecting channel, and a long timescale associated with the loss of fluid through evaporation. 
Initially, the two droplets rapidly adjust their volumes (on a timescale $t\lesssim 1$\,s) via a pressure-driven pumping flow and attain a stable configuration. Evaporation then causes the droplet pair to slowly step-through a sequence of quasi-stationary states determined by the instantaneous volume of the system. 
If the two droplets have the same contact area, $a_1 = a_2 = a$, and total volume $V_\mathrm{tot} \gtrsim 4\pi a^3/3$, then the droplets take on an asymmetric configuration with the larger droplet feeding the smaller droplet, so the larger droplet shrinks due to evaporative losses while the smaller droplet maintains approximately the same size and shape. Once  $V_\mathrm{tot} \lesssim 4\pi a^3/3$, the two droplets take on the same shape, the pumping flow between them reduces to zero, and they evaporate out in unison.
The evaporation-driven time-evolution of the droplet pair, from an asymmetric to a symmetric configuration, is underpinned by a symmetric pitchfork bifurcation (of quasi-static nature) in the case of droplets of equal contact area. 
Introducing pillars of unequal radii, and hence contact area, breaks the symmetry of the system and leads to an unfolding of the bifurcation with the distance between the connected and disconnected branches increasing with imperfection parameter $a_1/a_2$.
In experiments, jumping from the disconnected to the connected branch manifests as a sudden droplet-shape switching which changes the sign of the pressure difference between channel ends and causes the flow in the connecting channel to reverse. 

Our findings permit engineering control of flow reversal in capillary micropumps via platform design, which could be harnessed for bidirectional transport in point-of-care applications \cite{xing2011droplet}, and a framework for avoiding flow reversal in systems where a unidirectional flow is instead required throughout the entire evaporation process. Future research directions include examination of interactions between multiple evaporating droplets connected via a network of channels, and the influence of the vapour phase on the interaction of nearby droplets connected via a short channel.


We finish by drawing a parallel between the behaviour of interconnected droplets and that of two connected rubber balloons \cite{Muller2004}. 
In this canonical example of elasticity and stability \cite{Weinhaus1978,Meritt1978,Dreyer1982}, two balloons are inflated to the same size and connected to the two ends of a single pipe with a closed valve. Once the valve is opened the two balloons can exchange air and, if filled to an intermediate size, the initial configuration is unstable and the symmetry is broken instantly as air flows from one balloon to the other until a stable equilibrium is reached with an asymmetric configuration comprising one smaller and one larger balloon. The two balloon demo is analogous to our droplet experiments performed on equal pillars since, in both cases, multiplicity arises from nonlinear pressure-volume relations and equilibrium states can be achieved for both unequal and equal shapes of finite (i.e., non-zero) radii. 
However, in the balloon experiment, multiplicity emerges from the nonlinear pressure-volume relation that is typically a consequence of both geometric and constitutive (i.e., material) nonlinearities, while droplets dynamics are solely governed by a geometric nonlinearity between volume and contact angle. We postulate that a pair of connected deflating balloons would traverse a analogous bifurcation pathway as is exhibited by interconnected droplets as they evaporate. Similarly, our findings on unequal pillars suggest that the underlying symmetry of the two balloons could be broken by using balloons with different intrinsic radii of curvature, for example. 

Recently, the inflation of connected balloons has been proposed as a means of generating interactions between individual hysteretic elements (or `hysterons’) that can switch between states in a manner that depends on the path history of their states \cite{Hysterons}. Interacting mechanical hysterons can store and process information and therefore be programmed to function as state machines \cite{interactinghysterons}. Pushing further the analogy between our interconnected droplets and connected balloons, we propose that interconnected droplets are an example of interacting fluidic hysterons. Hence multiple interconnected droplets have the potential to exhibit coupled switching orders with evaporation-rate dependent pathways and memory effects, and could even behave as a (short-lived) state machine. Indeed, our results demonstrate that the transition pathways of these interacting fluidic hysterons can be programmed via careful engineering of system properties; in particular, through manipulation of droplet contact area, the relative humidity of the environment and the volatility of the droplets. \\


\noindent\textbf{Data Availability:} The reported experimental data are available in the SI.

\noindent\textbf{Funding:} CR acknowledges support from the China Scholarship Council (grant No.~202106230077). FB  acknowledges support from the Royal Society (URF/R1/211730 \& RF/ERE/210192).

\noindent\textbf{Conflicts of Interest:} The authors declare no conflict of interest and the funders had no role in the design of the study; in the collection, analyses, or interpretation of data; in the writing of the manuscript, or in the decision to publish the results.

\balance
\bibliographystyle{rsc}
\bibliography{drops}


\end{document}